\documentclass[twoside]{article}
\usepackage{fleqn,espcrc2}
\usepackage{graphicx}

\title{Effect of nonmagnetic impurities on stripes in high-$T_c$ cuprates}

\author{{T. Tohyama, M. Takahashi, and S. Maekawa}
\address{Institute for Materials Research, Tohoku University, Sendai, 980-8577, Japan.}}

\begin{document}

\begin{abstract}
We perform the numerically exact diagonalization study of the $t$-$J$ model with nonmagnetic impurities to clarify the relation between Zn impurities and the stripes.  By examining the hole-hole correlation function for a two-hole $\sqrt{18}$$\times$$\sqrt{18}$ cluster with a single impurity, we find that the impurity has a tendency to stabilize vertical charge stripes.  This tendency is caused by the gain of the kinetic energy of holes moving along the stripes that are formed avoiding the impurity.

$\;$\par
PACS: 74.20.Mn, 74.62.Dh, 71.10.-w

Keywords:  Stripe, Nonmagnetic impurity, $t$-$J$ model

\vspace{1pc}
\end{abstract}

\maketitle

Nonmagnetic Zn impurities substituted for Cu cause dramatic change of the electronic states of the high-$T_c$ cuprates, since Zn induces spatially inhomogeneous spin and charge distributions.  On the other hand, stripes observed in the cuprates have intrinsically inhomogeneous charge distribution \cite{Tranquada}.  Therefore, it is interesting to study the interplay of Zn impurities and stripes.  From experimental side, indications of the close relation between Zn impurities and stripes have been reported: The 1/8 hole-concentration anomaly of $T_c$, which is an indication of the stripes, is induced in Bi$_2$Sr$_2$CaCu$_2$O$_8$ and YBa$_2$Cu$_3$O$_{7-\delta}$ by Zn substitution \cite{Akoshima1,Akoshima2}.

In order to examine the relation between Zn impurities and stripes, we perform the numerically exact diagonalization study of the $t$-$J$ model with nonmagnetic impurities. By examining the hole-hole correlation functions for a small cluster with a single impurity, we find that the impurity has a tendency to stabilize vertical charge stripes.

We use a two-dimensional $t$-$J$ model with an impurity site.  The Hamiltonian reads
\begin{eqnarray}
H&=& J\sum\limits_{\left<i,j\right>,i,j\neq0}{{\bf S}_i}\cdot {\bf S}_j
    + J_{\rm imp}\sum\limits_{\left<0,j\right>}{{\bf S}_0}\cdot {\bf S}_j
     \nonumber \\
&&-t\sum\limits_{\left<i,j\right>,i,j\neq0,\sigma}
    \left( \tilde{c}_{i\sigma}^\dagger \tilde{c}_{j\sigma} +  \tilde{c}_{j\sigma}^\dagger \tilde{c}_{i\sigma}\right) \nonumber \\
&&-t_{\rm imp}\sum\limits_{\left<0,j\right>,\sigma}
    \left( \tilde{c}_{0\sigma}^\dagger \tilde{c}_{j\sigma} + \tilde{c}_{j\sigma}^\dagger \tilde{c}_{0\sigma} \right)\;,
\label{H}
\end{eqnarray}
where $\tilde{c}_{i\sigma }^\dagger$ is an electron creation operator at site $i$ with spin $\sigma$ with the constraint of no double occupancy, and the summation $\left< i,j \right>$ runs over all the nearest-neighbor pairs.  The impurity is located at site 0.  Here $t$ is the hopping parameter and $J$ is the exchange interaction.  $t_{\rm imp}$ ($J_{\rm imp}$) represents the hopping (exchange) strength of four bonds around the impurity site.  The case of $t_{\rm imp}$=$t$ and $J_{\rm imp}$=$J$ corresponds to the $t$-$J$ model, while the special case of $t_{\rm imp}$=$J_{\rm imp}$=0 describes the model with a vacant (nonmagnetic impurity) site.  We call the latter the Zn model hereafter.  We note that the paring properties of doped holes in the Zn model have been investigated using small clusters \cite{Poilblanc,Riera}.  In the present study, we use a $\sqrt{18}$$\times$$\sqrt{18}$ cluster with a single impurity and with two holes under periodic boundary condition (PBC) to clarify the relation between Zn impurities and stripes.  $J$ is taken to be 0.4$t$.

Let us start with the $t$-$J$ model without impurity.  It is controversial whether the $t$-$J$ model itself has the stripe-type ground state~\cite{SRWhite1,Hellberg}.
For small-size clusters, the stripe-like state is not stable in the ground state but appears as an excited state \cite{Hellberg}.  In the two-hole 18-site $t$-$J$ cluster, the ground state is a uniform state with the B$_1$ irreducible representation of $C_{4v}$ point group ($d_{x^2-y^2}$ symmetry).  By examining excited states, we identify one of excited states as the stripe-type state.  This state belongs to the $E$ representation having degeneracy of $p_x$ and $p_y$ symmetries.  We note that the degeneracy is simply lifted by introducing anisotropic terms that favor either the vertical or horizontal distribution of holes.  We have confirmed that the excited state identified as the stripe state lowers its energy under the presence of a potential that stabilizes vertical charge distribution and becomes the ground state when the potential exceeds a critical value as discussed in \cite{Tohyama}.

\begin{figure}{t}
\includegraphics[width=7.5cm]{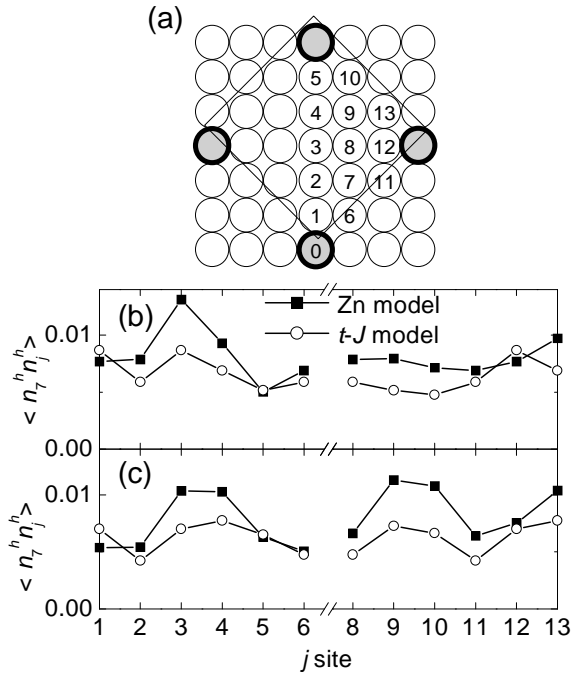}
\vspace{-27pt}
\caption{(a) $\sqrt{18}$$\times$$\sqrt{18}$ cluster with PBC.
The bold circles represent impurity sites that are repeated by PBC.  A part of sites are numbered.  (b) Hole-hole correlation $\left< n_7^h n_j^h \right>$ of the state with $d_{x2^-y^2}$ symmetry in a two-hole 18-site cluster.  $j$ runs over the sites numbered in (a).  Open circles and solid squares represent the correlations for the $t$-$J$ model and Zn model, respectively.
(c) The same as (b) but for the state with $p_x$ symmetry.
}
\label{fig1}
\end{figure}

To see the charge distribution in the ground state and the excited state, hole-hole correlation $\left< n_7^h n_j^h \right>$ is examined, where $\left <\cdots \right>$ is the expectation value for a given state, $n_j^h$ is the number operator of hole.  One of the two holes is fixed on the position 7 in Fig.~1(a).  The dependence of the correlation on the position $j$ of the other hole is shown in Fig.~1(b) and(c).
In the state with $d_{x^2-y^2}$ symmetry (the ground state), the positions $j$=1, 3, and 12 show maximum in the correlation, indicating that the probability that the holes are on diagonal sites is greater than nearest-neighbor sites \cite{SRWhite2}.  In contrast, the two holes in the state with $p_x$ symmetry obtained after lifting the degeneracy applying a very small anisotropic interaction have large probability to be found as a pair on sites with the distance of 2 and $\sqrt{5}$ (see the positions $j$=0, 4, 9, and 13).  It should be noticed that the correlation along the vertical pair, for instance, 7-9 is greater than that of a horizontal pair of 7-5 under PBC.  This means that the vertical distribution of the two holes is favorable in the $p_x$ state.  Hereafter we call the $p_x$ state the ``stripe'' state, while the $d_{x^2-y^2}$ state the ``uniform'' state.

The nonmagnetic Zn impurities simulated by setting $t_{\rm imp}$=$J_{\rm imp}$=0 induce spatially inhomogeneous charge distributions.  Figure~2 shows the difference of the hole density at site $j$ from the average value of hole density (2/18$\simeq$0.11).  It is found that the hole density near the Zn impurity (site 0) is reduced, while the enhancement of the density takes place at sites apart from the impurity.  The loss of the kinetic energy near the impurity induces such hole distribution \cite{Riera,Odashima}.  This behavior is seen in both the ``uniform'' and ``stripe'' states.

\begin{figure}
\vspace{5pt}
\includegraphics[width=7.5cm]{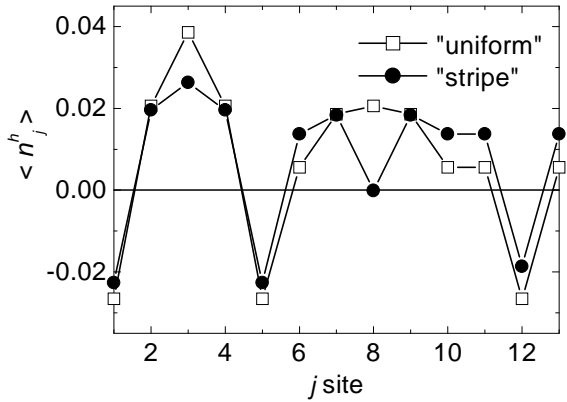}
\vspace{-27pt}
\caption{Difference of the hole density $\left< \Delta n_j^h \right>$ at site $j$ from the value for uniform hole distribution (2/18$\simeq$0.11).  The position of Zn impurities and site labeling are shown in Fig.~1(a).  Open squares and solid circles are for the state with $d_{x^2-y^2}$ symmetry (the ``uniform'' state) and with $p_x$ symmetry (the ``stripe'' state), respectively.}
\label{fig2}
\end{figure}

An important effect of the impurity is clearly seen when the energies of the two states are compared: The energy difference between the ``stripe'' and ``uniform'' states, $E_{\rm stripe}$$-$$E_{\rm uniform}$, decreases from 0.34$t$ to 0.15$t$ in the presence of the impurity.  This means that the ``stripe'' state is more sensitive to the impurity.  In other words, the Zn impurity has a tendency to stabilize the vertical charge stripes.  More interestingly, the ``stripe'' state could be the ground state by introducing long-range hopping terms whose magnitude is within a realistic parameter region \cite{Takahashi}.  An intuitive explanation about the sensitive behavior of the ``stripe'' state to the impurity is easily obtained through the examination of the hole-hole correlation.  From Fig.~1(b), it is found that the correlation $\left< n_7^h n_j^h \right>$ in the ``uniform'' state shows a prominent enhancement only at $j$=3 whose position is the farthest from the impurity.  Putting this result and other hole-hole correlations (not shown here) together, we can say that the holes are predominantly confined to the region surrounded by the impurities.  On the other hand, the ``stripe'' state exhibits enhancements of $\left< n_7^h n_j^h \right>$ at not only $j$=3 but also $j$=8 and 9.  Therefore, the holes in this state can get ability to gain the kinetic energy through the motion in the region between the impurities.  This contributes to lowering the energy of the ``stripe'' state under the presence of the nonmagnetic impurities.

In summary, we have perform the numerically exact diagonalization study of the $t$-$J$ model with nonmagnetic impurities to clarify the relation between Zn impurities and stripes.  By examining the hole-hole correlation functions for a small cluster with a single Zn impurity, we have found that the Zn impurity has a tendency to stabilize vertical charge stripes.  This comes from the gain of the kinetic energy trough the motion of holes along the stripes that are formed away from the impurity.

This work was supported by Priority-Areas Grants from the Ministry of Education, Science, Culture and Sport of Japan, CREST, and NEDO.  Computations were carried out in ISSP, University of Tokyo; IMR, Tohoku University; and Tohoku University.


\begin{thebibliography}{14}
\bibitem{Tranquada}J. M. Tranquada, B. J. Sternlieb, J. D. Axe, Y. Nakamura and S. Uchida S, Nature, London {\bf 375} (1995) 561.
\bibitem{Akoshima1} M. Akoshima, T. Noji, Y. Ono, Y. Koike, Phys. Rev. B {\bf 57} (1998) 7491.
\bibitem{Akoshima2} M. Akoshima, Y. Koike, I. Watanabe, K. Nagamine, Phys. Rev. B {\bf 62} (2000) 6761.
\bibitem{Poilblanc} D. Poilblanc, D. J. Scalapino, W. Hanke, Phys. Rev. Lett. {\bf 72} (1994) 884.
\bibitem{Riera} J. Riera, S. Koval, D. Poilblanc, F. Pantigny, Phys. Rev. B {\bf 54} (1996) 7441.
\bibitem{SRWhite1}S. R. White, D. J. Scalapino, Phys. Rev. Lett. {\bf 80} (1998) 1272; {\it ibid.} {\bf 81} (1998) 3227.
\bibitem{Hellberg}C. S. Hellberg, E. Monousakis, Phys. Rev. Lett. {\bf 83} (1999) 132.
\bibitem{Tohyama}T. Tohyama, S. Nagai, Y. Shibata, S. Maekawa, Phys. Rev. Lett. {\bf 82} (1999) 4910.
\bibitem{SRWhite2}S. R. White, D. J. Scalapino, Phys. Rev. B {\bf 55} (1997) 6504.
\bibitem{Odashima} S. Odashima, H. Matsumoto, Phys. Rev. B {\bf 56} (1997) 126.
\bibitem{Takahashi} M. Takahashi, T. Tohyama, S. Maekawa, unpublished.

\end{thebibliography}
\end{document}